\documentclass[aps,apl,superscriptaddress,twocolumn]{revtex4}
\usepackage{amsmath,bm}
\usepackage{graphicx,epsfig,braket,amssymb,amsfonts,eufrak,tabularx,multirow,amsmath,graphicx,color}

\newcommand{\bd}{\begin{displaymath}}
\newcommand{\ed}{\end{displaymath}}
\newcommand{\vv}[1]{{\bf #1}}
\newcommand{\bpm}{\begin{pmatrix}}
\newcommand{\epm}{\end{pmatrix}}
\newcommand{\nn}{\nonumber \\}

\begin{document}

\title{Topological waveguide based on metamaterials of magnetic solitons}

\author{Gyungchoon Go}
\email{gyungchoon@gmail.com}
\affiliation{Department of Materials Science and Engineering, Korea University, Seoul 02841, Korea}

\author{Ik-Sun Hong}
\affiliation{KU-KIST Graduate School of Converging Science and Technology, Korea University, Seoul 02841, Korea}

\author{Seo-Won Lee}
\affiliation{Department of Materials Science and Engineering, Korea University, Seoul 02841, Korea}

\author{Se Kwon Kim}
%\email{kimsek@missouri.edu}
\affiliation{Department of Physics and Astronomy, University of Missouri, Columbia, Missouri 65211, USA}

\author{Kyung-Jin Lee}
%\email{kj_lee@korea.ac.kr}
\affiliation{Department of Materials Science and Engineering, Korea University, Seoul 02841, Korea}
\affiliation{KU-KIST Graduate School of Converging Science and Technology, Korea University, Seoul 02841, Korea}

\begin{abstract}
We theoretically investigate coupled gyration modes of magnetic solitons whose distances to the nearest neighbors are staggered. In a one-dimensional bipartite lattice, analogous to the Su-Schrieffer-Heeger model, there is a mid-gap gyration mode bounded at the domain wall connecting topologically distinct two phases. As a technological application, we show that a one-dimensional domain-wall string in a two-dimensional soliton lattice can serve as a topological waveguide of magnetic excitations, which offers functionalities of a signal localization and a selective propagation of the frequency modes. Our result offers an alternative way to control of the magnetic excitation modes by using a magnetic metamaterial for future spintronic devices.
\end{abstract}

\maketitle

Topological properties embedded in band structures are one of the central themes in modern condensed matter physics.
In two-dimensional (2D) electron systems, representative examples supporting topologically protected edge states~\cite{Hatsugai1993} are Haldane model~\cite{Haldane1988} and Kane-Mele model~\cite{Kane2005}, which exhibit the quantum Hall and quantum spin Hall phases, respectively.
A classical example in one-dimensional (1D) topological systems is Su-Schrieffer-Heeger (SSH) model supporting a mid-gap bound state with fermion number one-half~\cite{Su1979, Jackiw1976}.
Inspired by the topological effects in electronic systems, numerous studies have been devoted to investigating topological properties in bosonic systems such as magnons~\cite{Katsura2010,Onose2010,Matsumoto2011}, phonons~\cite{Zhang2010, Zhang2015}, and their hybridized states~\cite{Takahashi2016, Zhang2019, Park2019, Go2019}.

Such topological effects of band structures can also be realized in artificially structured composites, called metamaterials, whose functionalities arise as the collective dynamics of local resonators~\cite{Ma2016}.
Analogues of topologically protected edge states in 2D systems have been proposed and experimentally observed in acoustic~\cite{Yang2015, Ma2016}, optical~\cite{Haldane2008, Wang2009, Cheng2016, Harari2018, Bandres2018}, magnetic~\cite{Shindou2013, Kim2017, Li2018a}, mechanical~\cite{Wang2015, Nash2015}, and electric circuit~\cite{Albert2015, Li2018b, Zhu2019} systems.
Moreover, the 1D SSH model has been realized in optical waveguides~\cite{Zeuner2015}, electric circuits~\cite{Lee2018a, Cai2019}, and magnetic spheres~\cite{Pi2018}.
An intriguing feature of the metamaterials is that the band structures and their topological properties can be manipulated by changing the crystal parameters.
This tunability of metamaterials is of crucial importance for widespread applications of topological properties in, for instance, reconfigurable logic devices~\cite{Ma2016,Cheng2016}.

Magnetic solitons such as magnetic vorticies and skyrmions are resonators whose dynamics exhibit gyroscopic motion~\cite{Guslienko2002, Mochizuki2012, Onose2012}.
Theoretical~\cite{Shibata2003, Shibata2004, Han2013} and experimental~\cite{Barman2010,Vogel2010,Sugimoto2011} results on the dynamics of coupled gyration modes of the magnetic solitons
provide potential application for a new type of information device~\cite{Han2013}.
Moreover, internal degree of freedoms of magnetic solitons such as polarity and chirality can offer efficient control of the functionalities of soliton-based metamaterials~\cite{Taniuchi}.
One of us has shown that the collective excitation of the magnetic solitons supports a chiral edge mode in honeycomb lattice~\cite{Kim2017}, which has been later confirmed by micromagnetic simulation~\cite{Li2018a}.
However, the SSH state in 1D system has not been realized for collective gyration modes of magnetic solitons.

In this paper, we first study a metamaterial composed of the magnetic soliton disks structured in a one-dimensional bipartite chain.
We show the existence of a mid-gap state bounded at a domain wall connecting topologically distinct two configurations, which is analogous to the electronic SSH model.
Then we consider a two-dimensional extension of our 1D magnetic SSH model, which is shown to be able to support a topological waveguide with selective propagation of frequency modes.

Here we consider a quasi-one-dimensional array of nanodisks containing magnetic vortices or skyrmions.
In general, the steady-state motion of topological solitons can be described by the dynamics of the center-of-mass ${\bf R}(t)$ and ${\vv m} = {\vv m} ({\vv r}- {\vv R}(t))$, where $\vv m$ is an unit vector along the direction of local magnetization.
The dissipationless magnetization dynamics of the coupled vortices/skyrmions is described by Thiele's equation \cite{Thiele}:
\begin{align}\label{GTE}
G \hat {\vv z} \times \frac{d{\vv U}_j}{dt^2}  + \vv F_j = 0,
\end{align}
where ${\vv U}_j = {\vv R}_j - {\vv R}^0_j$ is the displacement of the soliton from the equilibrium position ${\vv R}^0_j$,
$G = -4\pi M_s t_D Q /\gamma$ is the gyrotropic coefficient, $M_s$ is the saturation magnetization, $t_D$ is the thickness of the disk, and $\gamma$ is the gyromagnetic ratio.
Here, $Q = \frac{1}{4\pi} \int dx dy {\bf m} \cdot(\partial_x {\vv m} \times \partial_y {\vv m} )$ is the topological charge which characterizes the topological solitons.
The topological charge of the magnetic vortices and skyrmions are $Q= \pm 1/2$  and $\pm 1$, respectively.
${\vv F}_j = -{\partial W}/{\partial {\vv U}_j}$ is the conservative force from the potential energy
\begin{align}\label{eqW}
W = \sum_j \frac{K}{2} {\vv U}_j^2 + \sum_{j\neq k} \frac{U_{jk}}{2},
\end{align}
where $K>0$ is the spring constant and $\vv U_j \equiv (u_j, v_j)$ is the displacement vector.
Here, $U_{jk}$ is the interaction energy between two solitons:
\begin{align}\label{eqU}
U_{jk}(d_{jk}) = I_x(d_{jk})  u_j u_k - I_y(d_{jk}) v_j v_k,
\end{align}
where $d_{jk}(=|\vv R^0_j - \vv R^0_k|)$ is the distance between centers of two neighboring disks, and $I_x(d_{jk})$ and $I_y(d_{jk})$ are interaction parameters between two disks.
This system of coupled magnetic solitons has been studied both theoretically and experimentally~\cite{Shibata2003,Shibata2004,Vogel2010,Sugimoto2011}.
In particular, the values of the parameters in Eqs.~(\ref{eqW},~\ref{eqU}) have been experimentally measured and theoretically calculated for certain sizes of soliton disks, which will be used below for numerical calculations.

\begin{figure}[t]
\includegraphics[width=85mm]{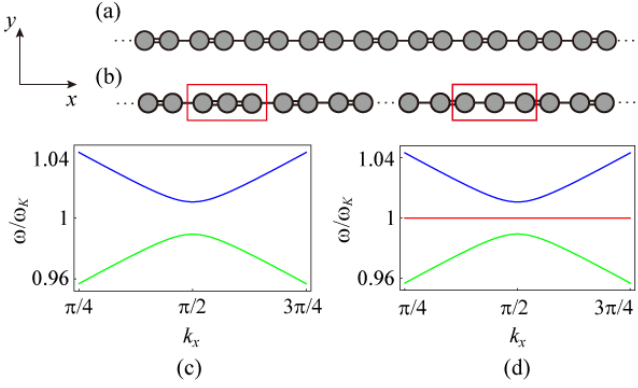}
\caption{A schmematic illustration a staggered 1D chain of magnetic nanodisks (a) and its dispersion (c) without domain wall defect.
A single (double) bond represents the longer (shorter) inter-disk distance.
A schmematic illustration a staggered 1D chain of magnetic nanodisks (b) and its dispersion (d) with a pair of domain wall and anti-domain wall defects.
For each defect, a mid-gap bound state appears at $\omega = \omega_K$ (red line). For calculation, we take the material parameters in Ref.~\cite{Li2018a}: $G = 1.09\times 10^{-14}$ Js$/$m$^2$, $K=3.74\times 10^{-4}$ J$/$m$^2$,
$I_x = -1.97\times10^{-6}$ J$/$m$^2$, $I_y = -1.19\times10^{-5}$ J$/$m$^2$, and $\Delta = 0.2$.} \label{fig:diskarray}
\end{figure}

Let us first consider the situation where the nearest-neighbor disk pairs are separated by a uniform distance.
Using the complex variable $\psi_j \equiv u_j + i v_j$, we write Eq.~\eqref{GTE} in a simplified form \cite{Kim2017,Li2018a}:
\begin{align}\label{eom}
i \dot{ \psi_j} = \omega_K \psi_j + \sum_{k\in \langle j \rangle} \left(\zeta \psi_k + \xi \psi^\ast_k\right),
\end{align}
where $\omega_K = K/G$ is the gyration frequency of an isolated soliton, $\zeta = (I_x-I_y)/G$ and $\xi = (I_x + I_y)/G$ are the reparameterized interactions.
In order to eliminate $\psi_k^*$, we expand the complex variable as
\begin{align}\label{cv}
\psi_j = \chi_j {\rm exp}({-i \omega_0 t}) + \eta_j {\rm exp}({i \omega_0 t}),
\end{align}
where $\chi_j (\eta_j$) is a counterclockwise (clockwise) gyration amplitude.
Substituting Eq.~\eqref{cv} into Eq.~\eqref{eom} and applying $|\chi_j|\gg|\eta_j|$ $(|\chi_j|\ll|\eta_j|)$ for counterclockwise (clockwise) soliton gyrations, we have
\begin{align}\label{ham}
i\dot{\psi_j} =\left(\omega_K - \frac{\xi^2}{\omega_K}\right) \psi_j + \zeta \sum_{k\in \langle j\rangle} \psi_k -\frac{\xi^2}{2\omega_K} \sum_{l\in\langle\langle j\rangle\rangle}\psi_l,
\end{align}
where $\langle\langle j\rangle\rangle$ represents second-neighbor sites of $j$.
The right-hand side of Eq.~\eqref{ham} contains zeroth order ($\omega_K$), first order ($\zeta$), and second order $(\xi)$ terms of the interdisk interactions.
Henceforth, we neglect the second order terms  which are smaller than others.
For 1D chain systems, we have
\begin{align}
i\dot{\psi_j} =\omega_K \psi_j + \zeta (\psi_{j+1} + \psi_{j-1}).
\end{align}
Taking the Fourier transformation, we obtain an eigenvalue equation, $i\dot{\Psi}({k},t) = H_k \Psi({k},t)$ with a momentum space Hamiltonian
\begin{align}\label{Hk0}
H_k &= \omega_K + 2\zeta \cos k \, ,
\end{align}
describing the single band of excitations.

Now, let us consider a staggered 1D chain of magnetic nanodisks [Fig.~\ref{fig:diskarray}(a)] with periodic boundary condition
which mimics the SSH system~\cite{Su1979}.
By introducing sub-lattice indices $A$ and $B$, we have
\begin{align}
i\dot{\psi}^A_{2m} &= \omega_K \psi^A_{2m} + \zeta_0(1+\Delta) \psi^B_{2m+1}\nn
&\hspace{16mm}+ \zeta_0(1-\Delta) \psi^B_{2m-1},\label{psidotA}\\
i\dot{\psi}^B_{2m+1} &= \omega_K \psi^B_{2m+1} + \zeta_0(1-\Delta) \psi^A_{2m+2}\nn
&\hspace{19.5mm}+ \zeta_0(1+\Delta) \psi^A_{2m}.\label{psidotB}
\end{align}
Here, $\Delta$, which can be either positive or negative, represents the staggeredness of the SSH system.
Taking the Fourier transformation yields
\begin{align}\label{Hk}
H_k &= \left(
  \begin{array}{cc}
    \omega_K  & 2\zeta_0 (\cos k_x - i \Delta \sin k_x) \\
    2\zeta_0 (\cos k_x + i \Delta \sin k_x) & \omega_K  \\
  \end{array}
\right)\nn
&= \vv n (k_x) \cdot {\boldsymbol \sigma},
\end{align}
where the basis of the Hamiltonian is $\Psi(k_x) = (\psi^A(k_x), \psi^B(k_x))^T$ and $\boldsymbol\sigma = (\sigma_x, \sigma_y)$ are Pauli matrices.
The dispersion relations of two modes are given by
\begin{align}
\omega_\pm = \omega_K \pm 2\zeta_0\sqrt{\cos^2 k_x + \Delta^2 \sin^2 k_x} \, . \label{eq:omega1d}
\end{align}
The staggeredness $\Delta$ induces a finite gap.
The topological number of the Hamiltonian~\eqref{Hk} can be computed in terms of a two-component unit vector $\hat {\vv n}(k_x) = {\vv n}(k_x)/|{\vv n}(k_x)| \equiv(\cos\theta_{k}, \sin\theta_{k})$,
as the integral:~\cite{Zak1989, Wu2012, Go2013}
\begin{align}\label{TN}
N = \frac{1}{2\pi} \int_{\rm BZ} dk_x \left(\frac{d\theta_k}{dk_x}\right) = {\rm sgn}(\Delta).
\end{align}
Eq.~\eqref{TN} implies that there are two topologically distinct phases which are represented by the sign of $\Delta$.

Expanding Eq.~\eqref{Hk} around ${k_x} = \pi/2$, which minimizes the band gap, we obtain a massive Dirac Hamiltonian
\begin{align}\label{Scheq1}
&H_k  = \left(
  \begin{array}{cc}
    \omega_K  & -2\zeta_0 ( {k_x} + i \Delta) \\
    -2\zeta_0 ( {k_x} - i \Delta) & \omega_K \\
  \end{array}
\right).
\end{align}
Diagonalizing Eq.~\eqref{Scheq1}, we obtain the eigenfrequencies with a band gap $\Delta$,
\begin{align}\label{omegapm}
\omega_\pm = \omega_K \pm 2\zeta_0\sqrt{{k_x}^2 +\Delta^2}.
\end{align}
Figure~\ref{fig:diskarray}(c) shows the dispersion relation of Eq.~\eqref{omegapm}.

\begin{figure}[t!]
\includegraphics[width=85mm]{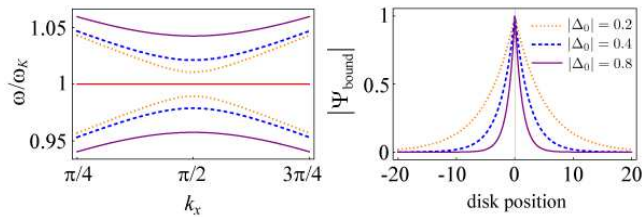}
\caption {Band structures (left) and localization of the bound state (right) for $\Delta_0 = 0.2$ (dotted), $\Delta_0 = 0.4$ (dashed), and $\Delta_0 = 0.8$ (solid).} \label{fig:loc}
\end{figure}

Because a topological bound state exists at the interface between the two topologically distinct phases, we consider the situation where the staggeredness $\Delta$ reverses its sign at $x=0$: $\Delta(x) =\Delta_0 {\rm{sgn}}(x)$.
In this case, a mid-gap bound state appears at $\omega = \omega_K$, without changing the bulk dispersions of upper and lower bands [see Fig.~\ref{fig:diskarray}(d)].
From Eq.~\eqref{Scheq1}, we read that the mid-gap bound state satisfies
\begin{align}
\left(
  \begin{array}{cc}
    0  &  i \partial_x - i \Delta(x) \\
    i \partial_x + i \Delta(x) & 0 \\
  \end{array}
\right)\Psi_{\rm bound} = 0,
\end{align}
which results in
\begin{align}\label{loc}
&\Psi_{\rm bound}(x) \sim \left(
         \begin{array}{c}
           0 \\
           e^{-\Delta_0 |x|} \\
         \end{array}
       \right) \quad  (\Delta_0 > 0),\nn
&\Psi_{\rm bound}(x) \sim\left(
         \begin{array}{c}
           e^{\Delta_0 |x|} \\
           0 \\
         \end{array}
       \right)\quad  (\Delta_0 < 0).
\end{align}
Eq.~\eqref{loc} shows that the bound state is exponentially localized at the domain wall.
This is a magnetic analogue of SSH system which possesses a soliton with half-electric charge~\cite{Su1979}.
Creation of the bound state is compensated by one-half of a state missing from the two bulk bands corresponding to $\omega = \omega_{\pm}$.
In Fig.~\ref{fig:loc}, we show the band structures and localization of the bound state for different values of $\Delta_0$.

\begin{figure}[t!]
\includegraphics[width=85mm]{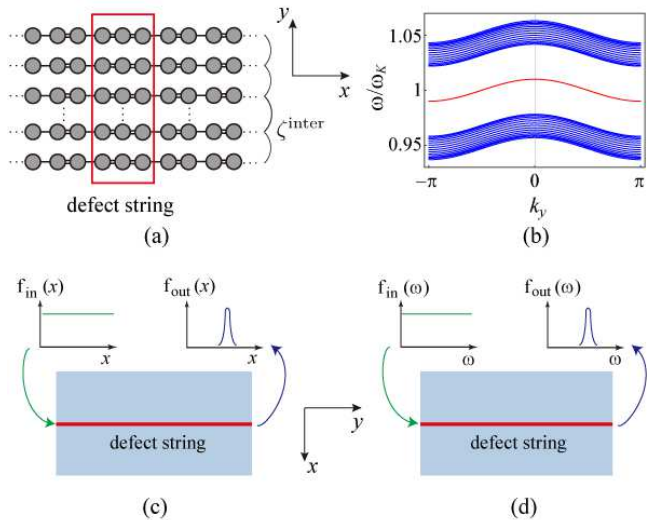}
\caption{(a) A schmematic illustration a two-dimensional extension of the 1D magnetic SSH model. (b) Bulk and bound state dispersions of the 2D model.
(c) Magnetic wave propagation in the topological waveguide supporting signal localization. (d) Magnetic wave propagation in the topological waveguide supporting selective propagation of frequency.}\label{fig:3}
\end{figure}

Now let us consider a 2D extension of our 1D magnetic SSH model, which will be shown to support a topologically protected waveguide of excitations below. The schematic illustration of the 2D lattice is shown in Fig.~\ref{fig:3}(a).
In the 2D lattice, the point-like defect in the 1D model is extended in $y$-direction and forms a domain wall string.
Given that the inter-chain interactions $\zeta^{\rm inter} \equiv (I^{\rm inter}_y - I^{\rm inter}_x)/G$ are uniform,
we have additional terms $\zeta^{\rm inter} (\psi^{A,B}_{y+1} + \psi^{A,B}_{y-1})$ to the 1D model equations [Eqs.~(\ref{psidotA},~\ref{psidotB})].
These terms yield an additional dispersion along $k_y$ direction, i.e., $\omega(k_y) = 2\zeta^{\rm inter} \cos k_y$.
The resultant 2D band dispersion is given by
\begin{align}\label{2Ddis}
\omega_{2D}(k_x, k_y) = \omega_{1D}(k_x) + 2\zeta^{\rm inter} \cos k_y,
\end{align}
where $\omega_{1D}$ is the band dispersion in the 1D lattice in Eq.~(\ref{eq:omega1d}).

In this 2D soliton lattice model, the topological mid-gap bound states are localized at the defect position and spatially connected in $y$ direction.
Therefore, magnetic excitations on the bound state propagate well along the defect string with a small spread in transverse ($x$) direction.
This propagation realizes a topological waveguide by using magnetic solitons with signal localization and selective propagation of frequency modes.
Fig.~\ref{fig:3}(c) and (d) show the schematic illustration of two functionalities of the magnetic soliton waveguide.
In Fig.~\ref{fig:3}(c), the incoming wave packet is a plane wave (i.e., uniform along $x$ direction) and has a frequency corresponding to the bound state.
Because this frequency mode can only propagate through the defect string, the outgoing wave packet is localized on the defect site.
In Fig.~\ref{fig:3}(d), the incoming wave packet on the defect site is a white signal having equal intensities for all frequencies.
However, most frequency modes on the defect site cannot propagate in $y$-direction, except for the bound state.
As a result, the outgoing wave packet on the defect site has a sharp peak at a frequency corresponding to the bound state.
From Eq.~\eqref{2Ddis}, we read that the frequency of bound state is determined by the gyrotropic frequency of a single magnetic soliton, which is tunable by external perturbations.
For example, in the presence of an effective magnetic field $H_{\rm eff}$ perpendicular to the disk plane, the gyrotropic frequency can be described as~\cite{Loubens2009, Ekomasov2017}:
\begin{align}
\omega \simeq \omega_K (1 + k H_{\rm eff}),
\end{align}
where $k$ is a proportionality constant. This suggests that the waveguide property can be manipulated by the external magnetic field or voltage-induced magnetic anisotropy change~\cite{Maruyama2009}.

To summarize, we have studied collective dynamics in a one-dimensional bipartite chain of the magnetic vortices or skyrmions.
In our magnetic system, the domain-wall topological defects are produced by changing the interdisk distances.
We have found that the topological defects induce the localized mid-gap states which are confined at the defect position.
Our finding on the 1D model is analogous to that of the SSH model in the electron system.
Compared to electronic SSH model which is hard to manipulate the domain walls, which are generated by the Peierl's instability, the topological manipulation is feasible in our magnetic SSH model.
As a technological application, we propose a two-dimensional extension of our 1D model, which supports a topological waveguide of magnetic excitations.
The topological waveguide provides not only a signal localization but also a selective propagation of the frequency modes.
Our work suggests that spintronics device based on magnetic metamaterials can offer a way for a precise control of the spin waves.

\hspace{10mm}

G. G. is supported by the National Research Foundation of Korea (NRF) (NRF-2019R1I1A1A01063594).
S.K.K. was supported by Young Investigator Grant (YIG) from Korean-American Scientists and Engineers Association (KSEA) and Research Council Grant URC-19-090 of the University of Missouri.
K.-J. L. acknowledges a support by the NRF (NRF-2017R1A2B2006119).

G. G. and I.-S. H. contributed equally to this work.

\end{document}